\documentstyle[12pt,epsf]{article}

\title{\large\bf 
Two-loop  renormalization  group profile of\\ 
the  standard  model and a new generation}
\author{Yu.~F.~Pirogov and O.~V.~Zenin\\[1 ex]  
{\normalsize\em Institute for High Energy Physics},\\
{\normalsize\em Protvino, Moscow Region, Russia}\\[0.5ex]
{\normalsize\em  Moscow  Institute  of  Physics  and  Technology,}\\
{\normalsize\em  Dolgoprudny,  Moscow  Region, Russia}}
\date{}

\begin{document}
\maketitle
\begin{abstract}
\footnotesize   
\noindent
The two-loop  renormalization group global profile of the Standard
Model (SM) in its full parameter space is investigated.  Restrictions
on
the Higgs boson mass as a function of a cutoff scale
are obtained from the stability of electroweak vacuum and from the
perturbative validity both in the Higgs and Yukawa sectors.
The cutoff equal to the Planck scale requires
the Higgs mass to be $M_H = (161.3 \pm 20.6)^{+4}_{-10}$ GeV and
$M_H\ge 140.7^{+10}_{-10}$ GeV, where the $M_H$ corridor is the
theoretical one and the errors are due to the top mass
uncertainty.  Modification of the two-loop global profile of
the SM extended by one new chiral generation is studied, and bounds
on the masses of the generation are derived. Under the precision
experiment
restriction $M_H\leq 200$~GeV, the fourth chiral generation,  taken 
alone, is excluded.  Nevertheless a pair of the chiral generations
constituting  the vector-like one  could exist.
\end{abstract}

\noindent
The renormalization group (RG) study of a field theory (for the
review see, e.g., Refs.~\cite{peterman,shirkov}) enables one
to understand in grosso the structure of the theory as a function of
a characteristic energy scale. Of special interest are the cases when
self-consistency of the theory is under danger of violation. They may
signal either the breakdown of the perturbative validity or/and the
onset of a ``new physics''.

There are two problems of the kind in the Standard Model (SM).
First, it encounters when some of the running couplings tend to blow
up at the finite scales violating thus the perturbativity. In the
case of the $\phi^4$ scalar theory,
the problem is known for a long time as the triviality problem. 
Second, the problem occurs when a running coupling leaves  the
physical region at some finite scale. In the SM, this happens when
the Higgs quartic effective coupling becomes negative, indicating the
absence of a ground state in the quantum theory. It is the so-called
electroweak vacuum stability problem. It is a real problem of the
quantum field theory because this phenomenon takes place in the realm
of the perturbative validity. In the framework of the SM, the light
Higgs bosons resulting in the unstable electroweak vacuum should be
forbidden. On the other hand, if this  happens some new
physics  beyond the SM will be required to stabilize the vacuum. 

The SM self-consistency study in the framework of the one-loop RG
and restrictions thereof on the SM heavy particles, the Higgs boson
and the top quark, was undertaken in
Refs.~\cite{cabibbo,beg,lindner1}. A generalization to
the two-loop level was given in Refs.~\cite{lindner2}--\cite{lee}.
The one-loop RG restrictions on a new heavy chiral family were
studied in Ref.~\cite{novikov}, and that on the vector-like family
were investigated in Ref.~\cite{zheng}. In Ref.~\cite{pirogov}
the RG study of the SM extended by the fourth chiral family was
generalized  to the two-loop level. This required a
generalization of the SM two-loop $\beta$ functions to the massive
neutrino case which was presented. As a by-product, the
two-loop RG global profile of the SM in its parameter space at all
conceivable scales was investigated. In particular the
self-consistency restrictions
on the Higgs mass were refined. More generally, the problem
of what principally new the fourth heavy chiral family brings in the
RG global profile of the SM  was considered.  A short summary of
these results is given in the report.

\subsection*{1. Renormalization group profile of the SM}

The SM two-loop $\beta$ functions in the $\overline{\rm MS}$
renormalization scheme are well-known in the literature
\cite{machacek}--\cite{ford2}  
(compact summaries can be found in Refs.\ \cite{barger,schrempp}). 
In what follows we put just the generic structure of the emerging
one- and two-loop RG differential equations. 
So let $g_{i}$, $ i=1,2,3$, $y_ f$, $\lambda$,
$v$ and $\mu$ be the SM gauge couplings, the Yukawa couplings for
fermions  $ f$, the  Higgs self-interaction 
coupling, the vacuum expectation value (VEV)
and the renormalization scale, respectively.
We neglected  for simplicity by the mixing of the Yukawa
couplings  and, thus, by  the CP violating phase.

The following essential features of the SM RG system 
are readily  ascertained. At one-loop order, one has a kind of the
three-level up-down hierarchy among the SM couplings, so that
the  differential equations for $g_i$, $y_f$ and $\lambda$
disentangle. One can first find  $g_{i}(\mu)$, then
insert them into  $\beta_{y_{f}}^{(1)}$ and find $y_{f}(\mu)$, and
finally put  $g_{i}(\mu)$, $y_{f}(\mu)$ into the equation for
$\lambda$ and integrate it.
The solution to the equation for $v$ is determined completely by
those for the first three equations, both in one and two loops.

In two loops, the  RG equations partially entangle with each other
due to a down-up feedback to the neighbour level: from $\lambda$ to
$y_{f}$ and from $y_{f}$ to $g_{i}$. But there is no direct influence
of $\lambda$ on $g_{i}$. It emerges only in three loops. Hence to
completely entangle the RG system  one needs the three-loop SM
$\beta$ functions, which are unknown at present.  Thus we have to
restrict ourselves to the two-loop order. On the other hand, the two-
and higher-loop contributions to $\beta$ functions, even the sign
including, are known to depend in a multi-coupling theory on the
renormalization scheme \cite{shirkov}.
Hence the physical meaning of the running couplings becomes
ambiguous, and it is impossible to improve the perturbative RG
analysis of the SM in the scheme-independent way beyond one loop.

We integrated the SM RG system  numerically for $\mu\ge M_Z$ by the
first-order Runge-Cutta method with the initial conditions at the 
scale $M_{Z}$ taken as $\alpha_{1}(M_Z)=0.0102$,
$\alpha_{2}(M_Z)=0.0338$ and $\alpha_{3}(M_Z)=0.123$ 
in accordance with  $\alpha(M_Z)=1/127.90$ and 
$\sin^{2}\theta_{W}(M_Z)=0.2315$~\cite{rpp}. 
Our normalizations of the gauge couplings are as follows:
$g_{1}=(5/3)^{1/2}g'$, $g_2\equiv g$ and $g_3\equiv g_S$, with $g'$,
$g$ and $g_S$  being the conventional SM couplings. We choose also
the relations $m_{f}=y_{f}v$  and $m_{H}=\lambda^{1/2} v$  as the
definition of normalization for the Higgs and Yukawa couplings, with
$v=(\sqrt{2}\,G_F)^{-1/2}=246.22$ GeV being the Higgs VEV. Because
the evolution of $v(\mu)$ is gauge dependent we use in what follows
only the gauge independent observable $v\equiv v(M_Z)$.
Besides we used  at $\mu=M_Z$ the one-loop  matching condition
for the physical  and  running fermion masses  $M_f$ and
$m_f(\mu)\equiv y_f(\mu)\,v$, respectively. We  have got at
$M_H=150$~GeV:
$m_\tau(M_Z)=1.764$ GeV, $m_b(M_Z)=(4.47\pm 0.50)$ GeV and
$m_t(M_Z)=(171.8^{+4.6}_{-4.7})$ GeV. The last two values correspond
in turn to the physical bottom and top masses $M_b=(4.5\pm 0.5)$
GeV ~\cite{rpp} 
and  $M_{t}=(175 \pm 5)$ GeV~\cite{top}, respectively. Only errors in
the top mass are left as the main source of the subsequent
uncertainties. 

As a field theory, SM is legitimate to be pulled to its inner
ultimate limits. This may help to  understand better its structure in
the physically reasonable region  $\mu<M_{Pl}$, $M_{Pl}\simeq
10^{19}$
GeV, which is to be considered more seriously. So all the 
numerical results were obtained at all allowed $\mu$ with the exact
two-loop $\beta$ functions. Let us discuss the results in turn for
the gauge, Yukawa and Higgs sectors of the~SM.

\subsubsection*{({\bf\em i}) Gauge  sector}

\begin{figure}[htbp]

\epsfysize=70mm\epsfbox[-50 0 400 300]{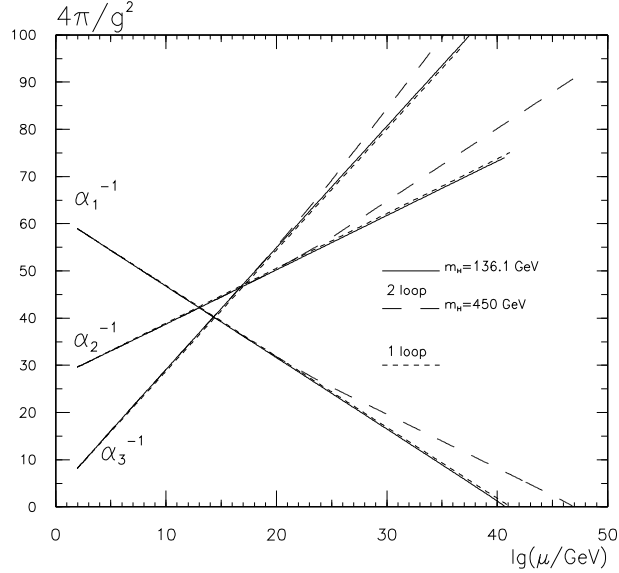}
\caption{\footnotesize
Running of the inverse gauge couplings squared
$\alpha_i^{-1}$,~$i=1,2,3$.
Number of generations is $n_g=3$. Represented Higgs masses are those
corresponding to the typical heavy Higgs and to the lower 
critical Higgs curve  shown in bold in Fig.~3.}
\end{figure}

\noindent
Fig.~1  shows the running with $\mu$ of the inverse  gauge
couplings squared. It is seen that the  coupling $g_1$ develops a
pole singularity  at $\Lambda_{g_1}$, $\log\Lambda_{g_1}\simeq 41$.
Validity of the perturbation theory in $g_1$ restricts
$\alpha_1\le4\pi$ and hence $\log\mu\le40$, which is in the
logarithmic scale twice as large as the Planck scale. We should
assume this restriction on the physical grounds in what follows.
Nevertheless, taken at its face value the
two-loop RG has the meaning by itself. So, to  understand better the
mathematical structure of its solutions we extend them up to the
singularity point $\Lambda_{g_1}$. The actual influence of
$y_f(\lambda(\mu))$ on 
$g_1$ in two loops is somewhat sizable only for the heavy Higgs.  It
diminishes the slope of $g_1(\mu)$ at $\mu$ beyond the Planck scale,
where $y_f$ are large, and shifts the singularity position
$\Lambda_{g_1}^{(2)}$ upwards to $\log\Lambda_{g_1}^{(2)}=47$ for the 
heavy Higgs ($m_H(M_Z)=450$ GeV). 
The critical value ~$m_H(M_Z)=136.1$ GeV corresponds to the maximal
lower bound of the vacuum stability. 

\subsubsection*{{(\bf\em ii}) Yukawa  sector}

\begin{figure}[htbp]
{\epsfysize=80mm \epsfbox[-100 0 400 390]{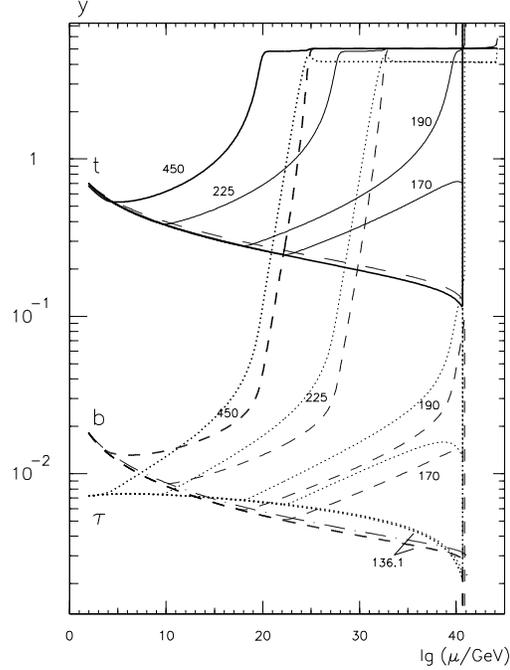}}
\caption{\footnotesize
Running of the third family Yukawa couplings ($n_g=3$).
The falling down curves shown in bold correspond to the lower
critical Higgs mass. The thin lines, close to the latter bold ones,
correspond to the one-loop approximation.}  
\end{figure}

\noindent
Fig.~2 presents the evolution of the Yukawa couplings $y_f$ 
for the third family SM fermions: $t$, $b$  quarks, and $\tau$
lepton. In one loop,  all the $y_f$ are falling down with $\mu$ and
lie in the weak coupling regime. To be more precise, the 
one-loop trajectory for the $\tau$ lepton is mildly convex, so that
it intersects with the curve for the $b$ quark near the GUT scale. 
But in two loops the behaviour changes
drastically. An approximate UV stable  fixed point appears at
$y_t^{(\mbox{\scriptsize UV})}\simeq 5.2$.
In the real world, this critical value is approached from below both
for $t$, $b$ quarks and for $\tau$ lepton, the faster the heavier
Higgs boson is. (see Fig.~1).   
Hence for the sufficiently heavy Higgs, $m_H\ge 200$ GeV,  all the
third family fermions
would fall into the strong coupling regime at sufficiently high
$\mu$. This would make the  third family fermions much more alike at
the high scales than at the electroweak one. In practice, prior to
$M_{Pl}$ the strong coupling develops only for $t$ quark when Higgs
is rather heavy, $m_H\ge 450$ GeV.
Because from the combined LEP data on the precision experiments  it
follows that $M_H\le 200$~GeV at 95\% C.L.~\cite{q98}, one may
conclude that the Yukawa sector of the SM is weakly coupled along all
the physically reasonable region of $\mu$, $\mu\le M_{Pl}$.

\begin{figure}[htbp]
\epsfysize=80mm \epsfbox[-100 0 400 420]{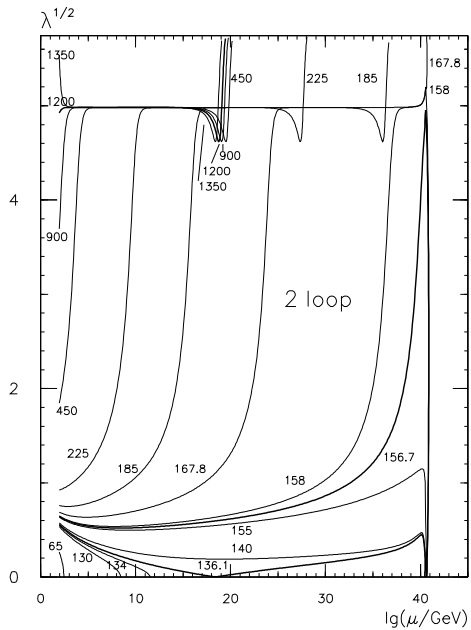}
\caption{\footnotesize 
The two-loop running of the Higgs quartic coupling ($n_g=3$). The
critical curves are shown in bold.}
\end{figure}

\begin{figure}[htbp]
{\epsfysize=80mm\epsfbox[-135 20 400 400]{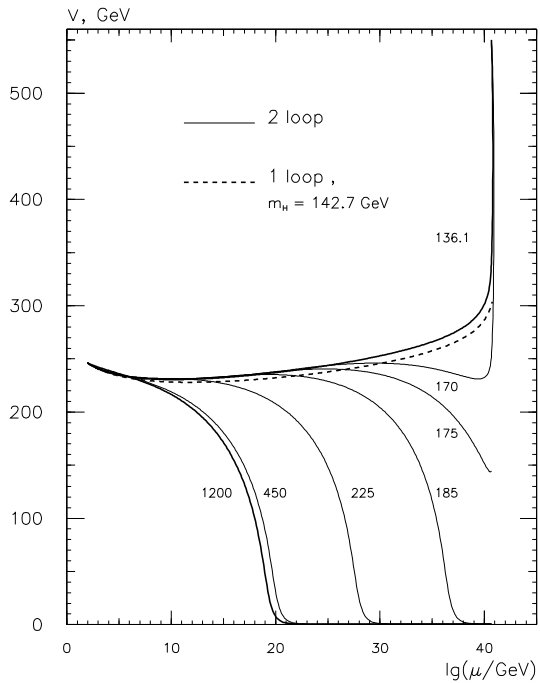}}
\caption{\footnotesize
Running of the  Higgs VEV ($n_g=3$) in the 't~Hooft--Landau gauge.}
\end{figure}

\subsubsection*{({\bf\em iii}) Higgs  sector }

Fig.~3 presents running of the Higgs quartic coupling.
In two loops, there are three critical curves shown in bold.
First of all, there appears an approximate UV stable fixed
point at $\lambda^{1/2}_{\mbox{\scriptsize UV}} \simeq 4.93$ produced
by the compensation of the one- and two-loop terms: $\lambda^2$ and
$\lambda^3$. It corresponds
to boundary value of the Higgs mass $m_{H~max}^{(2)}(M_Z)=1200$ GeV,
at and above which the theory is definitely strongly coupled. 
The boundary Higgs mass for the vacuum instability  in two
loops is $m_{H~min}^{(2)}(M_Z)=136.1$ GeV. The third critical value
$m_{H~inter}^{(2)}(M_Z)\simeq 156.7$ GeV borders the region with the
potentially 
strongly coupled Higgs from the one with the weakly coupled Higgs.
Note that theory with $m_{H~min}^{(2)}<m_H(M_Z)<m_{H~inter}^{(2)}$ is
consistent in two loops up to the ultimate scale
$\mu=\Lambda_{g_1}^{(2)}$.
For completeness, we present in Fig.~4 the plot for $v(\mu)$  in the
't~Hooft--Landau gauge both in one and two loops. It is seen that the
electroweak symmetry never restores prior to the Plank scale.  

Finally, one can impose the requirement of the SM self-consistency
up to some cutoff scale $\Lambda$. In other terms, the theory should
be neither strongly coupled nor unstable at $\mu\le\Lambda$.
In one loop, this means that the $\lambda$ singularity
position fulfills the requirement $\Lambda_{\lambda}^{(1)} \ge
\Lambda$, and simultaneously one has $\mu\mid_{\lambda=0} \ge
\Lambda$. In two loops, we should choose as a criterion for the onset
of the strong
coupling regime the requirements $\beta_\lambda^{(2)}/
\beta_\lambda^{(1)}|_\Lambda$ and $ \beta_\lambda^{(3)}/
\beta_\lambda^{(2)}|_\Lambda<1 $ which could guarantee the
perturbativity and the scheme independence. 
In neglect by all the couplings but $\lambda$ this would  mean that
$\lambda^{1/2}\leq 2$, in particular $m_H(M_Z)\leq 500$ GeV,
the
restriction we retain for the whole SM.
Not knowing $\beta_t^{(3)}$ we restricted ourselves just by the
requirement  that $\beta_t^{(2)}/ \beta_t^{(1)}|_\Lambda\ll 1 $ which
is definitely fulfilled at $y_t\leq 2< y_t^{(UV)}$.

\begin{figure}[htbp]
{\epsfxsize=80mm \epsfbox[-150 0 400 300]{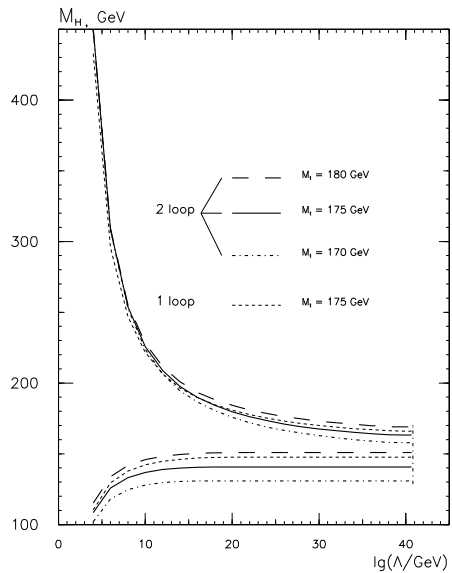}}
\caption{\footnotesize 
The SM one- and two-loop  self-consistency plot ($n_g=3$): the
allowed Higgs mass vs.\ the cutoff scale~$\Lambda$.}
\end{figure}

The resulting one- and two-loop restrictions are drawn in Fig.~5.
Here
the transition was made from the $\overline{\rm MS}$
value $m_H(M_Z)$ to the physical Higgs mass $M_H$.
The sensitivity of the allowed region of the Higgs mass 
to the uncertainty of the top quark mass
is also indicated. Strictly speaking, allowed is the region between
the most upper and the most lower curves. This means that for
$\Lambda=M_{Pl}$ the legitimate Higgs mass  is $M_H = (161.3 \pm
20.6)^{+4}_{-10}$ GeV. One gets also  the lower bound  $M_H\ge
140.7\pm 10$ GeV at such a cutoff.  The allowed Higgs mass
region is much wider for a lower cutoff scale $\Lambda$.

\subsection*{2. Restrictions on the fourth chiral family }

We considered the minimum SM extension by means of the additional
heavy fermion families. If alone, the fourth family should have with
necessity the same chirality pattern as the three light families.
This is to be required
to avoid the potential problem of  the large direct mass mixing
for the fourth family with the light ones.

What concerns the fifth family, there are two possibilities: either
it has the same chirality as the four previous families, or it is a
mirror one
(or to state it differently, it is charge conjugate with respect to
the rest of the families). In the first case, the analysis repeats
itself just with more parameters. In the second case, the large
direct mass terms could be
introduced for the pair of the fourth and fifth families, in addition
to Yukawa couplings. This proliferates enormously the number of free
parameters and makes the general analysis impossible. On the other
hand, if one chooses a mass independent renormalization scheme, say 
$\overline{\rm MS}$, the net influence of the direct mass terms on
the evolution of the SM parameters will be just in the threshold
effects. Barring them, this case, which may likewise be attributed to
one vector-like family, is technically equivalent to the case with
two chiral families.

For these reasons we restricted ourselves by one new chiral family.
In order to conform with experimental value for the number 
of light neutrinos ($n_{\nu}=3$), we should also add the right-handed 
neutrinos $\nu_R$ (at least for the fourth family) and the proper
Yukawa couplings for them. The right-handed neutrinos may possess the
explicit Majorana mass as well, so that the physical neutrino masses
may be quite different from their Yukawa counterparts. Because in the
mass independent
renormalization the explicit mass terms are important only in the
threshold effects, we disregard them in what follows. We generalized
the two-loop SM $\beta$ functions of Ref.~\cite{machacek} to the case
with the neutrino Yukawa couplings~\cite{pirogov} . 
For practical calculations with the fourth family we neglected by
the light neutrino Yukawa couplings.

At present, there are no theoretical hints on
the existence (or v.v.) of the fourth (and the higher) family.
Nevertheless, one can extract some restrictions on the corresponding
fermion masses. They are twofold, direct and indirect ones, being in
a sense complementary to each other. The first group gives bounds on
the common mass scale of the fourth family, the second one restricts
the splitting of the masses inside the family.

The existing direct experimental bounds on the masses of the fourth
fami\-ly quarks $t_4$ and $b_4$ depend  somewhat on the assumptions
about their decays. If the lightest of the quarks, say $b_4$, is
stable enough to leave the 
detector, the limit on its mass is $M_4 \ge 140$ GeV \cite{cdf}.
On the other hand, for unstable quarks, decaying inside the detector,
the limit can be estimated from the CDF and D0 searches for the top
quark~\cite{top} to be  about $M_t$. What concerns the neutral and
charged  leptons of the fourth family, $\nu_4$  and $e_4$, it follows
from LEP searches that $M_{\nu_4} \ge 59$ GeV and $M_{e_4} \ge 90$
GeV  at 95\% C.L.~\cite{rpp,lep}.

The indirect restrictions can be extracted from the precision
electroweak data, and they are related to the absence of decoupling
with respect to the heavy chiral fermions. This results in the
quadratically growing dependence of
the electroweak radiative corrections on the heavy fermion
masses. To avoid such a large radiative corrections,
as the precision data require, 
the members of a heavy fermion doublet should be highly degenerate.
Namely, one should have for the quarks $t_4$ and $b_4$ that 
$(M_{t_4}^2 - M_{b_4}^2)/M_Z^2 \le 1$, and similarly for the leptons
$\nu_4$, $e_4$. One important peculiarity of the
vector-like family is the decoupling with respect to the explicit
mass term, at the Yukawa couplings being finite. Hence unlike the
chiral family, there is no need here for the high degeneracy in the 
Yukawa couplings to suppress the large
radiative corrections.

To reduce the number of free parameters we assume in what follows
that $m_{t_4} = m_{b_4} = m_Q$ and $m_{\nu_4} = m_{e_4} = m_L$.
As representative, we considered two cases: $m_L/m_Q = 1/2$ and 1, 
with the common mass $m_Q$ of the heavy quarks given by the fourth
family scale $m_4$. It follows that both these typical cases do not
contradict to the direct experimental bounds if $m_4 \ge 180$ GeV.
Our results for the case $m_4 = 200$ GeV which we
consider as more  realistic and $m_L/m_Q = 1/2$ are presented in
Figs.~6--10.

\begin{figure}[htbp]
{\epsfysize=70mm \epsfbox[-100 0 400 330]{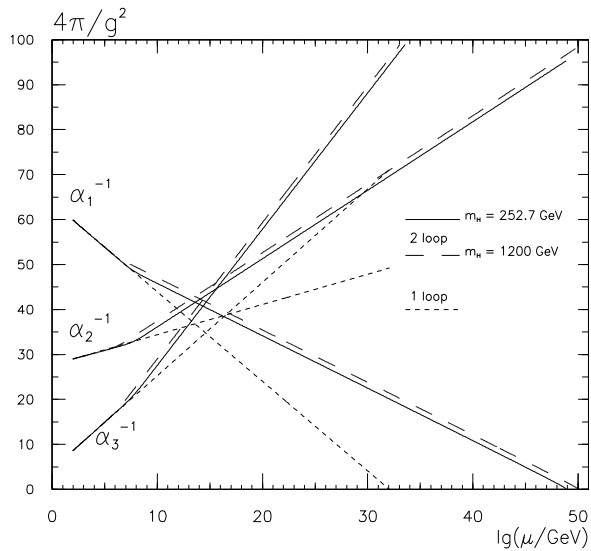}}
\caption{\footnotesize
Running of the gauge couplings ($n_g = 4$). The fourth family mass
scale is $m_4 = 200$ GeV and $m_L/m_Q = 1/2$.}
\end{figure}

Fig.~6 shows the evolution of $\alpha_i^{-1}$ with $\mu$. 
It may be noted that the GUT triangle
shrinks and shifts towards somewhat lower scale but the conceivable
gauge unification takes place beyond the
region of perturbativity both in Yukawa and Higgs sectors. 
It is seen that the two-loop contributions manifest themselves at
rather low scales, $\mu = (10^7 - 10^8)$ GeV. They are governed by
the onset of the strong coupling regime in the Yukawa sector at such
a $\mu$ (see Fig.~7).
Accordingly, the perturbatively consistent region of $\mu$ in the
Higgs sector shrinks to the same values (see Fig.~8).

\begin{figure}[htbp]
{\epsfysize=70mm \epsfbox[-150 -20 400 330]{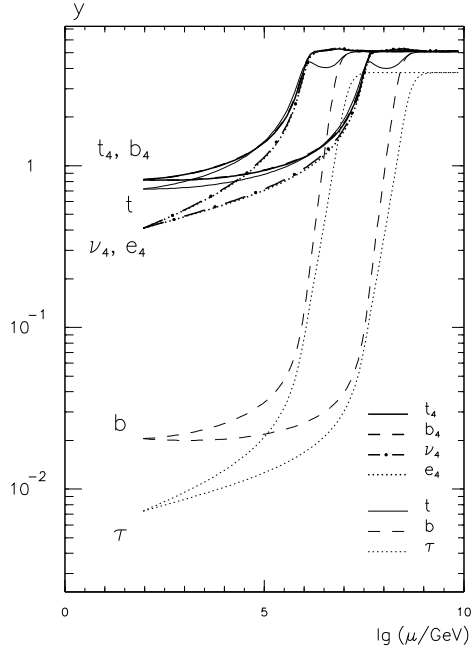}}
\caption{\footnotesize
Two-loop running of the Yukawa couplings ($n_g = 4$) for the third
and fourth families at $m_4 = 200$ GeV and $m_L/m_Q = 1/2$.
The upper and lower curves correspond to the Higgs masses,
respectively, for the upper and lower Higgs critical curves shown in
bold in Fig.~8.}
\end{figure}

Applying now the same criteria of self-consistency as in the case of
the minimal SM we get the allowed values
of $m_H(M_Z)$ depending on the cutoff scale $\Lambda$ (Fig.~9). 
The sensitivity to the shift in the mass $m_4$ is also indicated.
Finally, Fig.~10 presents the two-loop allowed
region in the $m_4$--$M_H$ plane. The dependence on $\Delta M_t$ is
faint, and it is not shown.

Let us summarize the differences in the RG global profiles of the
SM with three and four generations. For three generations with 
the experimentally known masses,
the Yukawa sector is weakly coupled in the one-loop approximation.
Prior to the Planck scale,
the strong coupling may appear in one loop only
in the Higgs self-interactions for the sufficiently heavy Higgs.
It drives strong coupling for the Yukawa sector as well, but only
through two loops. As a result, this influence is reduced, and the
Yukawa sector stays weakly coupled up to the Planck scale for all
experimentally preferred values of the Higgs mass, $M_H \le 200$ GeV.
Validity of the perturbative SM up to the Planck scale,
the Yukawa sector including, as well as the vacuum stability require
the Higgs mass to be $M_H = (161.3 \pm 20.6)^{+4}_{-10}$ GeV and
$M_H\ge 140.7^{+10}_{-10}$ GeV. Here the $M_H$ corridor is the
theoretical one and the errors are produced by the top
mass uncertainty. The allowed Higgs
mass region is wider for a lower cutoff scale~$\Lambda$.

\begin{figure}[htbp]
{\epsfysize=60mm \epsfbox[-150 0 400 300]{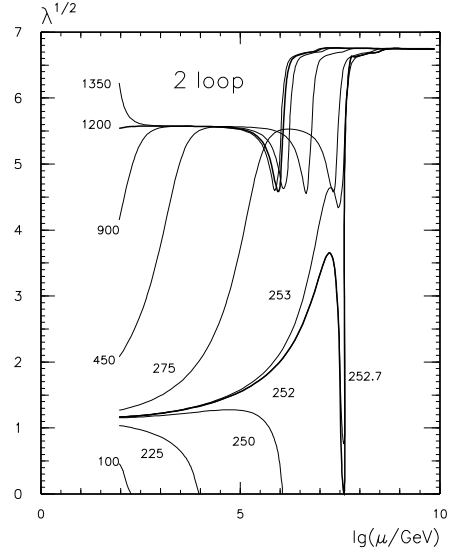}}
\caption{\footnotesize
Two-loop running of the Higgs quartic coupling ($n_g=4$) at   
$m_4 = 200$ GeV and $m_L/m_Q = 1/2$.}
\end{figure}

\begin{figure}[htbp]
{\epsfxsize=90mm \epsfbox[-260 0 400 300]{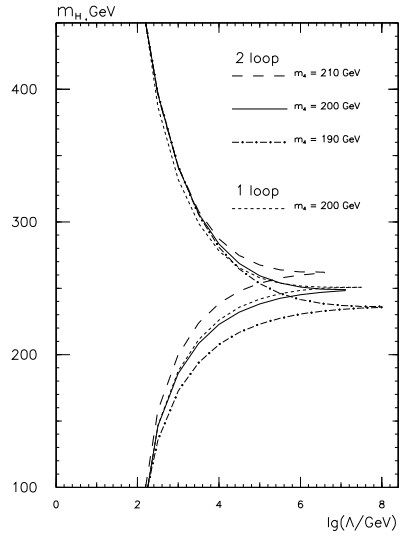}}
\caption{\footnotesize
One- an two-loop self-consistency plots ($n_g=4$): the allowed Higgs
mass vs.\ the cutoff scale $\Lambda$  at $m_4 = 200$ GeV and 
$m_L/m_Q = 1/2$.}
\end{figure}

The inclusion of the fourth heavy chiral family qualitatively changes
the mode of the SM realization.
With the addition of the family, the strong coupling is driven 
in one loop by the Yukawa interactions. It transmits to the Higgs
self-interactions at the one-loop order, too. 
Hence the strong coupling develops in both these sectors
in parallel, and their coupling constants blow up at sufficiently low
scales. 
As a result, the requirement of self-consistency of the perturbative
SM as an underlying theory up to the Planck or GUT scale excludes
the fourth chiral family. But as an effective theory, the SM allows
the heavy chiral family with the mass up to 250 GeV depending on the
Higgs mass and the cutoff scale.  Under precision experiment
restriction $M_H\leq 200$~GeV, the fourth chiral family,  taken
alone, is excluded. Though  a pair of the chiral families
constituting  the vector-like one  could still exist.

\begin{figure}[htbp]
{\epsfysize=80mm \epsfbox[-100 0 400 380]{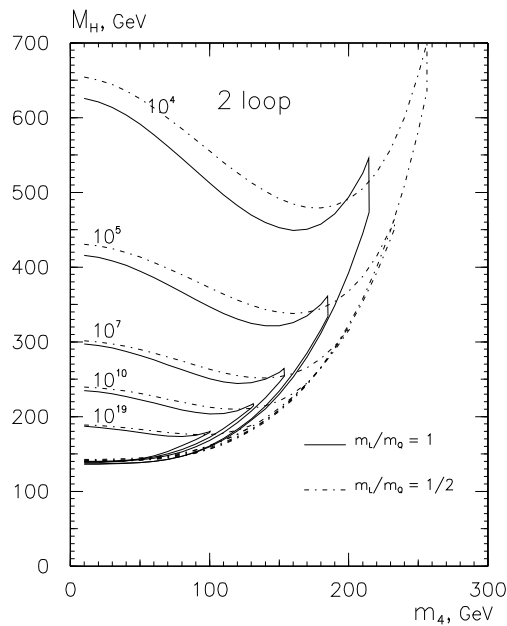}}
\caption{\footnotesize
Two-loop self-consistency plot under the $y\leq 1.5$  restriction on
the Yukawa couplings ($n_g=4$).}  
\end{figure} 

\paragraph{Acknowledgements}
This work was supported by the RFBR under grant No.~96--02--18122.
One of us (Yu.F.P.) is grateful fo the Organizing Committee of the
Seminar for support.

\end{document}